# Chemical Potential Shift in Doped Mott-insulators for Energy Storage Applications

Chutchawan Jaisuk[1], Tanawat Sawasdee[1], Warakorn Jindata[1], Thanundon Kongnok[1], Sirichok Jungthawan[1], Atsushi Fujimori[2,3] and Worawat Meevasana[1,a]

**AFFILIATIONS:**

[1]School of Physics, Institute of Science, Suranaree University of Technology, Nakhon Ratchasima 30000, Thailand

[2]Department of Physics, University of Tokyo, Tokyo 113-0033, Japan

[3]Department of Physics and Center for Quantum Science and Technology, National Tsing Hua University, Hsinchu 30013, Taiwan

[a]Authors to whom correspondence should be addressed: worawat@g.sut.ac.th

**ABSTRACT:**

This work explores the unique character of strongly correlated systems, specifically Mott-insulators, in the context of battery electrode materials. The study investigates the correlation between the proposed chemical potential evolution and charge storage performance in transition metal oxide-based electrodes. The hypothesis suggests that doping a Mott insulator reduces the Hubbard Coulomb interaction, which could slow down the chemical shift and result in enhanced charge storage capabilities compared to classic band insulators. The results support the hypothesis through a systematic comparison of selected transition metal oxide-based electrodes (Cu, Mn, Co, and Fe oxide electrodes). Furthermore, a toy model is employed to investigate the shift in chemical potential with doping-dependent U using DFT+U calculation, aiming to visualize the chemical potential evolution in Mott-insulators relevant to their application as battery electrodes. This study provides valuable insights into how strongly correlated materials, especially Mott-insulators, contribute to the advancement of energy storage technologies.

## I. INTRODUCTION

Battery technology is currently a major focus of research worldwide, driven by the escalating demand for energy storage solutions to power portable electronics, electric vehicles, and renewable energy systems. The demand for increased energy density remains a fundamental challenge in battery research with developing advanced materials and novel architectures to enhance energy density in order to satisfy specific energy requirements. Particularly, boosting the energy density of batteries is essential for enhancing the performance of mobile electronic devices. A lithium-ion battery (LIB) is a rechargeable battery belonging to the broader category of batteries, where lithium ions undergo movement from the anode electrode to the cathode electrode during discharge and reverse during the charging process [1]. Owing to its high energy density and rechargeability, the LIB has become the predominant battery type in portable consumer electronics and electric vehicles [2-3]. The inherent character of electrode materials significantly influences the overall performance of batteries. Currently, the cathode materials that have been commercialized are primarily composed of transition metal oxides, including the classic $LiCoO_2$ (LCO), $LiNiO_2$ (LNO), $LiMn_2O_4$ (LMO) and ternary metal oxides such as $LiNi_xMn_yCo_zO_2$ or NMC [4-5]. Specifically, it was suggested that during the charging process, $LiCoO_2$ undergoes the Mott transition while transforming into $Li_{1-x}CoO_2$ [6]. In the practical battery industry, transition metal oxide materials emerge as the predominant choice for commercially available battery electrodes. Fundamentally, it is a challenge to find out the intrinsic nature that makes the group of transition metal oxides outstanding as battery electrodes. Generally, the charge storage mechanism in transition metal oxide (TMO) based electrodes can be generalized as $M_xO_y + 2yLi^+ + 2ye^- \leftrightarrow xM + yLi_2O$, where M represents a 3d transition metal such as Cu, Mn, Co, Fe, etc. Transition metal oxide (TMO)-based electrode materials are considered promising candidates due to their capability to exhibit remarkably high specific capacities. However, the elusive nature of the underlying mechanism that enables extraordinary high capacity in these materials remains a topic of debate [7].

While TMOs have been of much interest for energy storage applications, TMOs have also been of interest in condensed matter physics as strongly correlated systems. Specifically, the insulating character of many strongly correlated materials can be effectively described by the Mott-Hubbard picture [8]. The strongly correlated effect is responsible for the opening of the Mott-gap, resulting in insulating behavior. Doped Mott-insulators form a fascinating class of materials exhibiting a range of intriguing phenomena arising from the interplay between strong electron-electron interactions and charge carrier doping. This interplay alters the electronic and magnetic properties of the parent Mott-insulating state. The discovery of high-$T_c$ superconducting materials has sparked interest in doped strongly correlated materials. More precisely, the parent compounds of high-$T_c$ are known to be Mott-insulators [9-10]. This has led to more investigation of doped Mott-insulators. In this work, we will start looking into TMOs from the point of view of condensed matter physics, then focusing on the possible effect of Mott-gap closing upon doping, and finally relating this to the resulting chemical potential shift which could enhance the battery energy capacity.

Fundamentally, the chemical potential shifting is a parameter used for determining the evolution in electronic structure upon doping. In this article, we report the possibilities of chemical potential evolution upon doping the Mott-insulator; the theoretically microscopic study of this question is indeed very complex and under much debate. We look at the empirical effect and experimental measurements. Regardless of the microscopic origin, in term of negative electronic compressibility (NEC), the negative thermodynamic density of states $K_e =$ $(1/n^2)(\partial n/\partial \mu) < 0$, where n is the carrier density and µ the chemical potential, can be described as a counterintuitive decrease in chemical potential upon increasing electron density [11-12]. The NEC effect can be experimentally observed by using photoemission spectroscopy (PES) measurements [13-17]. It is obvious that the system with slower chemical potential shifting upon carrier doping exhibits great capability in electronic compressibility [16]. With this slower chemical potential shifting, the host can store more charge carriers while maintaining the potential (V); this suggests implication for energy storage applications [17-18]. Recently, with the interest in the field of energy storage applications, the electronic structure and chemical potential shift has been used to study for the understanding and optimizing the performance of batteries, particularly in enhancing energy density. For example, the electronic band structure and density functional theory (DFT) calculations were used to provide the information about the electronic structure of electrode materials which helps in understanding the electrochemical properties of the electrode materials [19-23]. Introducing dopant atoms into the crystal structure can also modify the electronic structure of electrode materials, which can improve their conductivity and enhance energy density. For example, the enhancement of capacitance in Li-ion batteries has been studied through the introduction of a dopant Ni atom into the $V_2O_5$ electrode. This introduces low-valence $Ni^{2+}$ states that occurred in the electronic structure of the host $V_2O_5$, leading to a 28% increase of capacitance in the Ni-doped $V_2O_5$ electrode when compared to the undoped $V_2O_5$ electrode [24].

We initiate our exploration by examining the scenarios of chemical potential shift in doped strongly correlated materials based on experimental observations to highlight the unusual phenomena associated with chemical potential shift. In previous research, Fujimori and his colleagues conducted thorough experimental investigations aimed at deducing the chemical potential shift from core-level photoemission in strongly correlated electron systems. Their work was focused on understanding the behavior of electrons in materials, particularly concerning the concept of chemical potential shift in correlated electron systems. The samples primarily consist of transition metal oxides, particularly the parent compounds of high-$T_c$ superconductors. A notable observation is the pinning character of the chemical potential in $La_{1-x}Sr_xTiO_{3+y/2}$, $La_{2-x}Sr_xCuO_4$, and $La_{2-x}Sr_xNiO_{4+y}$ within the region of low doping concentration via cation substitution [25]. This behavior is suggestively linked to the formation of stripes or microscopic charge ordering [26]. Furthermore, the character of chemical pinning is experimentally observed in doped manganite, specifically $Pr_{1-x}Ca_xMnO_3$ (PCMO). In particular, the pinning of the chemical potential in PCMO is attributed to a particularly charge-ordered state over the wide hole concentration region between x=0.3 and x=0.75 in the electronic phase diagram [27].

For rigid bands model, the electronic structure remains inflexible against carrier doping. An increase in electron filling directly leads to an increase in chemical potential. Surprisingly, however, doping a Mott-insulator results in a decrease in the band gap. Consequently, the chemical potential can decrease with an increase in electron concentration. Mathematically, the chemical potential evolution $\Delta\mu/\Delta n$ will be negative which is referred to as the NEC effect. An experimental study using angle-resolved photoemission spectroscopy (ARPES) on doped strongly correlated material $(Sr_{1-x}La_x)_3Ir_2O_7$ revealed that a decrease in chemical potential occurs as the La content increases, particularly with an increasing electron doping concentration [14]. It was specifically observed that the Mott-gap is suppressed in $(Sr_{1-x}La_x)_3Ir_2O_7$ as electron doping increases. The appearance of a smaller energy gap or the tendency for the Mott-gap to close in $(Sr_{1-x}La_x)_3Ir_2O_7$ causes a negative $\Delta\mu/\Delta n$, indicating an interesting electronic response in the system. Moreover, this smaller-gap behavior has also been observed in Rb-doped $Ca_3Cu_2O_4Cl_2$

Mott-insulator measured by ARPES [28]. Specifically, the Mott-gap undergoes a rapid collapse as spectral weight transfers from the charge transfer band into the band gap region, ultimately emerging in the upper Hubbard band. Though the changes within the Mott-gap area are complex, we can generally understand it as the Mott-gap closing scenario. This finding confirms a breakdown of the rigid band picture in describing scientific observations. In other words, it can be generalized that an increase in doping concentration weakens the interaction, resulting in a reduction in the correlation gap.

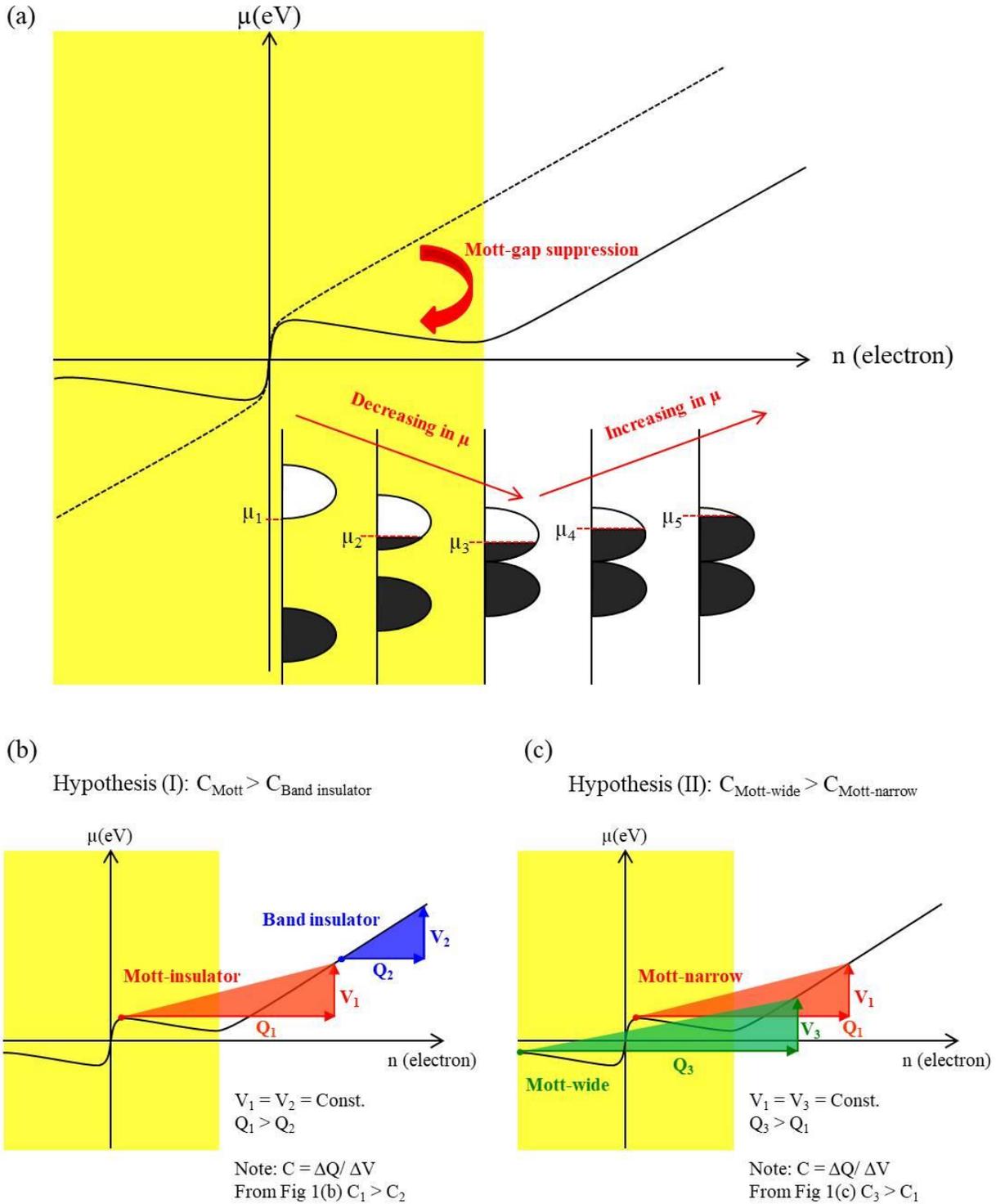

**FIG. 1.** (a) The schematic diagram illustrates the phenomenon of Mott-gap suppression, which transforms the chemical potential into a slowly increasing scenario. The solid line and dash line represent the chemical potential of the Mott-insulator and conventional rigid band, respectively.

The yellow area is referred to as the region of Mott-gap suppression effect. The hypotheses in (b) and (c) represent the ability to store electrical charge, which is related to the utilization of Mott-gap suppression. Defining the capacitance C=ΔQ/ΔV, hypothesis (I) presumes that the capacitance of a Mott-insulator should exceed that of a band insulator. Complementary, hypothesis (II) postulates that more utilization in Mott-gap suppression effect will results in greater charge storage capability.

Please note that the depiction of the chemical potential evolution of a Mott-insulator in Figure 1, illustrating the greater potential for charge storage of Mott-insulator compared to a band insulator, represents a simple scenario of chemical potential evolution. From the perspective of Mott-gap closing, the Mott-gap is expected to be continuously suppressed as the Mott-insulator is doped. Typically, the generalized scenarios for the evolution of chemical potential in a Mott-insulator are simplified into three categories, as shown in Figure 2. For more details, the slowly changing chemical potential (Figure 2(a)) presents the effect of the Mott-gap suppression upon charge doping, which slightly slows the rate of change in chemical potential. The chemical potential pinning scenario (Figure 2(b)) illustrates the unchanging chemical potential as the Mott-gap is continuously suppressed. The pinning of the chemical potential upon charge doping effectively delays changes in the chemical potential compared to those in the conventional rigid band filling scenario. Negative electronic compressibility (Figure 2(c)), where the chemical potential decreases as electron filling increases, occurs when the rate of Mott-gap closing is significantly faster than the rate of chemical potential rising due to electron filling. All three situations concerning the evolution of chemical potential in doped Mott-insulators fundamentally endorse the enhancement of charge capacity in Mott-insulators for battery electrodes.

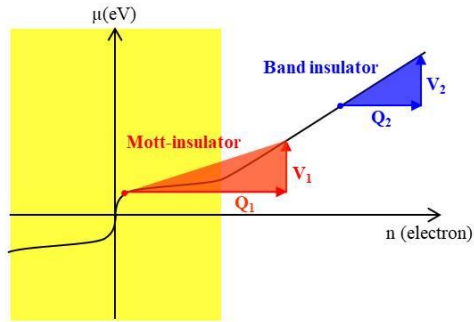

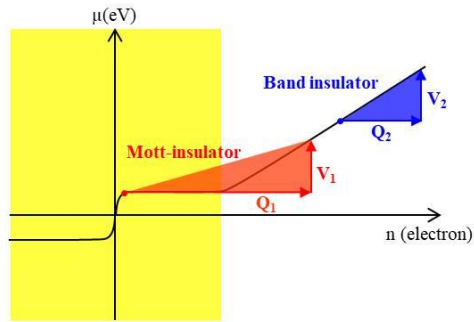

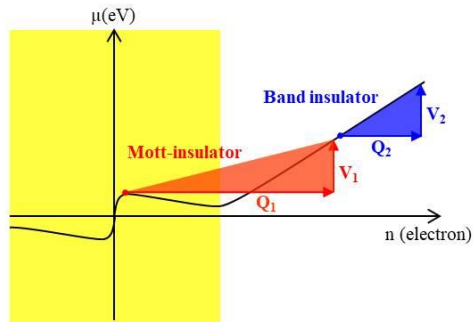

**FIG. 2.** Possible cases of chemical potential evolution of Mott-insulator that can enhance the performance of battery electrode including (a) slowly changing in chemical potential (b) the chemical potential pinning and (c) the negative electronic compressibility.

## II. HYPOTHESIS about Mott-Gap Closing Effect

Fundamentally, it has been found that many strongly correlated materials are utilized as electrode materials in batteries. The capability of storing the charge carrier of the electrode is the primary parameter used to determine the performance of the battery electrode. In the following, we describe the charge storage performance of strongly correlated materials, in particular, transition metal oxide based on the literature of the charge/discharge profile, including Cu, Mn, Co, and Fe oxide electrodes.

Specifically, we propose that Mott-gap closing upon doping would be the case that makes the chemical potential increase slower rather than normal. Therefore, this effect yields more stored charge capability while maintaining a slower voltage change compared to a rigid band scenario as shown in Figure 1. In classification, our hypothesis can be generalized into two points, as shown in Figure 1(b) and 1(c): (I) the capacitance of the Mott-state electrode should be larger than that of the band insulator electrode, and (II) the capacitance of wide Mott-sweeping (Mott-wide) should be larger than that of narrow Mott-sweeping (Mott-narrow). In our assumption, we define the terms "wide" and "narrow" Mott-sweeping based on the oxidation state of the transition metal atom in the Mott-insulator. During battery charging, or when electrons are added to the Mott-insulator electrode, the oxidation state of the transition metal decreases as more electrons are doped until the oxidation state approaches zero. From this perspective, a higher oxidation state in the transition metal atom represents a wide Mott-sweeping. We will demonstrate the consistency of the hypothesis by comparing the capacitance in TMO-based electrodes. The hypothesis (I) is supported by Cu and Mn oxide based electrodes and the hypothesis (II) is supported by Mn, Co and Fe oxide based electrodes.

## III. RESULTS

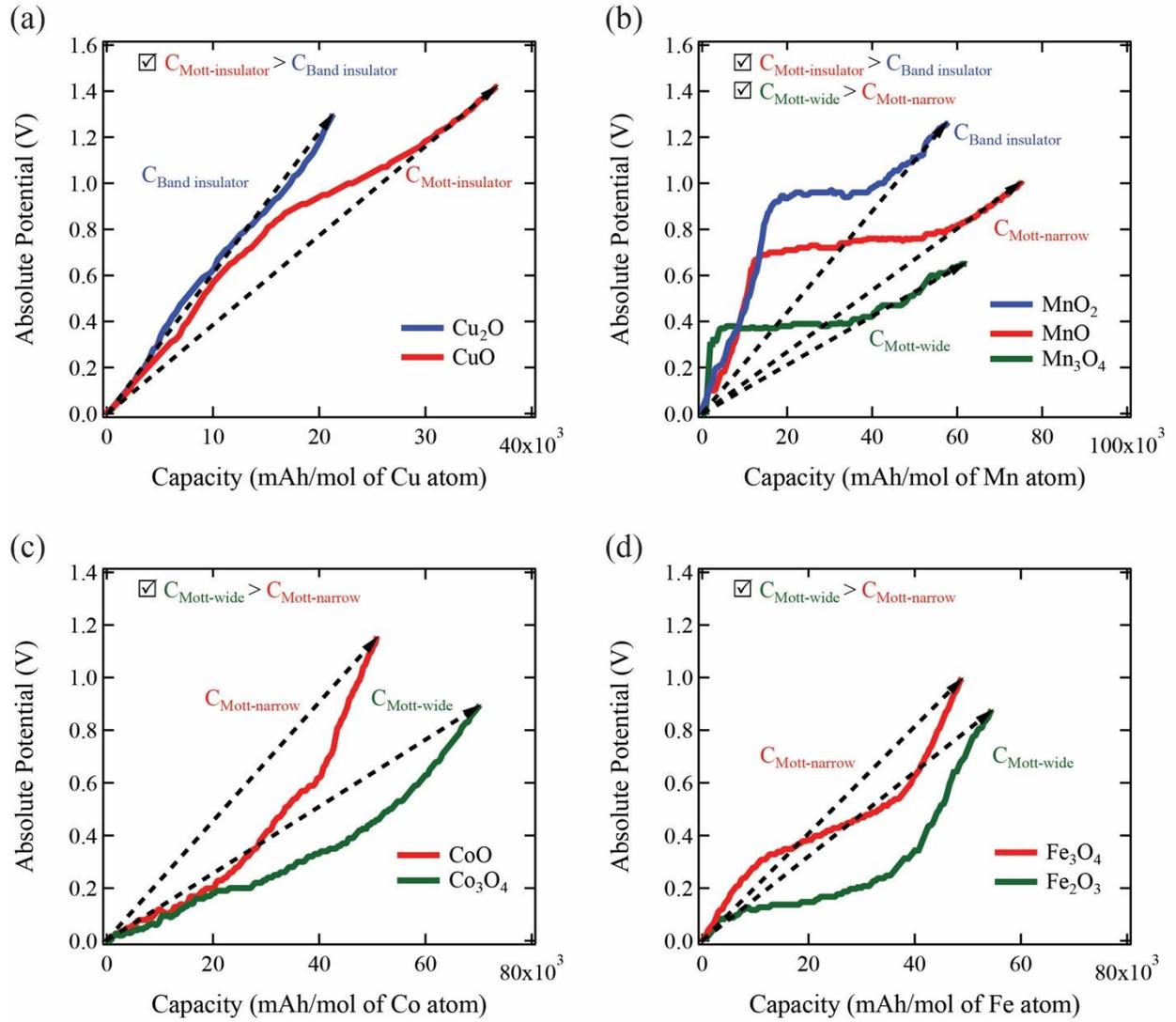

**FIG. 3.** Four panels illustrate the discharge profiles of (a) $Cu_2O$ and $CuO$, (b) $MnO_2$, $MnO$, and $Mn_3O_4$, (c) $CoO$ and $Co_3O_4$, and (d) $Fe_3O_4$ and $Fe_2O_3$. The arrow-dot line is introduced to distinguish the global slope of each discharge line. The electrode capacitance can be straightforwardly interpreted as 1/slope.

To verify the hypothesis given in Figure 1, we explore the charge storage performance of selected TMOs-based battery electrodes including Cu, Mn, Co, and Fe oxide electrodes based on the re-analysis from the experimentally reported data. To avoid the bias from our own measurements, we analyzed the results from available literatures [29-34]. The summarized results are illustrated in Figure 3, plotting the potential (V) versus capacity in mAh/mol of transition metal atom. As shown in Figure 3, the selected TMOs-based electrodes are compared in category based on type of transition metal composition.

**Copper-Based Electrodes (CuO, $Cu_2O$)**

In the field of energy storage, Cu-based oxide is one of the well-known transition metal oxides used as electrodes, including $Cu_2O$ and CuO. The electronic structure of $Cu_2O$ and CuO can be simply explained by the electronic configuration of electrons in the d-shell of the Cu transition metal atom. Typically, the Cu atom has an $[Ar]3d^{10}4s^1$ electronic configuration. In the case of $Cu_2O$, the rise of chemical bonding results in the $d^{10}$ closed shell occupation. Therefore, the insulating nature of $Cu_2O$ can be understood by the general picture of the fully filled d band. It has been experimentally found that the band gap of $Cu_2O$ is about 2.1 eV. Turning to CuO, the electronic configuration of the Cu atom exhibits a partial d shell occupation ($3d^9$). It has been investigated that the insulating nature of CuO is due to the effect of the electron-correlation effect in the open-shell d bands [35]. The Mott-gap of CuO is about 1.4 eV. To distinguish the electrochemical performance of $Cu_2O$ and CuO, the literature data about the capacity of $Cu_2O$ and CuO are compared here [29-30]. To create a new analysis, the chosen capacitance as a ratio of 70% of the charge capacity and a certain active potential will be used to estimate the charge storage capacity for the comparison. From the discharge profiles of CuO and $Cu_2O$, it can be mathematically found that the charge capacity of CuO (Mott-insulator) is 1.58 times that of $Cu_2O$ (band insulator). Alternatively, we provide the modified discharge profiles of $Cu_2O$ and CuO as shown in Figure 3(a). The inversion of the slope of the arrow-dot line in Figure 3 globally represents the capacity of each electrode. It is obvious that the capacitance of CuO (Mott-insulator) is greater than that of $Cu_2O$ (band insulator). This result satisfies our first hypothesis in Figure 1(b), which highlights the outstanding properties of the Mott-state as compared to the band insulator.

**Manganese-Based Electrodes (MnO, $Mn_3O_4$, $MnO_2$)**

The electrodes composed of manganese oxide also exhibit significant promise for utilization in battery applications. The familiar manganese oxide forms include MnO, $MnO_2$, and $Mn_3O_4$. MnO is a recognized example of a Mott-insulator. It is regarded as the representative of the perfectly half-filling Mott-insulator with Mn ($3d^5$) or $Mn^{2+}$ configuration. $Mn_3O_4$ is a semiconductor type with an average $Mn^{+2.67}$, or it can be imagined as a nearly Mott-state. $MnO_2$ is a semiconductor with a partially occupied d shell ($3d^3$ configurations or $Mn^{4+}$), and it can be imagined that $MnO_2$ is a nearly band-insulator type. In 2014, J. Yue et al. systematically synthesized and studied the electrochemical performances of manganese oxides [31]. Based on their experimental discharge profiles, the capacitance of MnO, $Mn_3O_4$, and $MnO_2$ can be mathematically determined for comparison. From Figure 3(b), it has been found that the highest charge storage capacitance among manganese oxides is that of $Mn_3O_4$. Specifically, the greater capacitance of $Mn_3O_4$ than that of MnO supports the second hypothesis that wide Mott-sweeping is greater than narrow Mott-sweeping. In addition, the capacitance of MnO (Mott-insulator) is greater than that of $MnO_2$ (band insulator), which satisfies the first hypothesis.

**Cobalt-Based Electrodes ($CoO$, $Co_3O_4$)**

Cobalt oxides have been recognized as potential electrode materials suitable for battery applications. The common forms of cobalt oxide used in electrochemical applications are referred to as CoO and $Co_3O_4$. The reversible electrochemical process exhibited by cobalt oxide electrodes within lithium-ion batteries (LIBs) can be represented as follows: $CoO + 2Li \leftrightarrow Li_2O + Co$ and $Co_3O_4 + 8Li \leftrightarrow 4Li_2O + 3Co$, for CoO and $Co_3O_4$ electrodes, respectively. Fundamentally, CoO is considered the classical Mott or charge transfer insulator with a band gap arising from the correlation effect. In terms of $Co_3O_4$, one can simply consider $Co_3O_4$ as the nearly Mott-state. We would like to refer to the research work done by Sun et al. in 2017 involving the electrochemical study of CoO and $Co_3O_4$ for the charge performance comparison [32]. From the discharge profiles of CoO and $Co_3O_4$ in Figure 3(c), we can find that $Co_3O_4$ (wide Mott-sweeping) has 1.43 times higher charge capacity compared to that of CoO (narrow Mott-sweeping). This result is in full agreement with the second hypothesis.

**Iron-Based Electrodes ($Fe_2O_3$, $Fe_3O_4$)**

Iron oxides have gained significant interest in being employed as battery electrode materials in lithium-ion batteries due to their abundance and low process cost. The particular forms of iron oxide utilized in electrochemical conversion reactions are $Fe_2O_3$ and $Fe_3O_4$. In terms of electronic structure, $Fe_2O_3$ is specifically considered a Mott-insulator. In the case of $Fe_3O_4$, one can generally view it as a closely near-Mott-state situation. In this review, we use the discharge profile of $Fe_2O_3$ (Mott-insulator) from the research study by Zhang et al., 2013, and of $Fe_3O_4$ by Pan et al., 2018, for making a comparison between $Fe_2O_3$ and $Fe_3O_4$ in terms of charge storage capability [33-34]. From Figure 3(d), $Fe_2O_3$ (wide Mott-sweeping) obviously delivers a higher charge capacity, around 1.23 times that of $Fe_3O_4$ (narrow Mott-sweeping). This situation complements the second hypothesis.

**Toy Model for Chemical Potential Shift with Doping-Dependent U**

Many experimental studies reported in the literature have focused on how the chemical potential evolves when the Mott-insulator is doped. The pinning of chemical potential was experimental observed by core level photoemission measurement [25, 27]. Moreover, scanning tunneling microscopy measurements offer further insights into the introduction of in-gap states upon doping charge into a Mott-insulator and lead to the pinning of the chemical potential, as demonstrated in prior studies [36-38]. Additionally, the negative compressibility, in which the change in chemical potential is opposite compared to the ordinary change in the rigid band model, arising from the suppression of the Mott-gap, was experimentally discovered [14-15]. The discussion on how the chemical evolves upon doping is considered a controversial topic. In this study, we will simply use the empirical effect in which the Hubbard coulomb interaction U continuously decrease to find the effect on the chemical potential of doped Mott-insulator.

Among various strongly correlated materials, nickel oxide (NiO) is widely recognized as the classic Mott-Hubbard representation. Specifically, NiO is classified within the category of charge transfer insulators [39-40]. Particularly, the band gap can be formed even in the presence of an odd number of electrons due to strongly correlated electron interaction. The Mott-gap of NiO has been accurately determined through experimental techniques, specifically employing Photoemission and Bremsstrahlung-Isochromat-Spectroscopy (BIS). The measured value is determined to be 4.3 eV [41]. In this study, NiO is chosen as the strongly correlated material for a toy model investigation to demonstrate the evolution of the chemical potential upon charge carrier doping under doping-dependent U.

The computational investigation employs density functional theory (DFT) calculations. The electronic structure of NiO was obtained using self-consistent DFT simulations with the projector augmented wave (PAW) method as implemented in the Vienna Ab initio Simulation Package (VASP) [42-43]. The calculation was performed in the framework of the DFT+U approach to include the correlation effect in the strongly correlated system. Specifically, the calculation utilizes the simplified DFT+U approach introduced by Dudarev et al. [44]. In practice, the initial effective Hubbard term ($U_{eff}$) is chosen at 7.0 eV, which is relatively close to reported data [45]. The cutoff energy was set at 250 eV, and a $\Gamma$-centered 4 x 4 x 4 Monkhorst-Pack k-mesh was used for the Brillouin zone integrations. Since the Mott-gap should be suppressed as electrons are continuously added, we simply employ decay function as a trial function for simplicity to represent the decreasing Hubbard Coulomb potential term as shown in Figure 4(b). The desired data obtained from the DFT calculation, representing the chemical potential evolution upon adding electrons, is measured in terms of the difference between the Fermi level ($E_F$) and the valence band maximum ($E_{VBM}$), particularly $E_F - E_{VBM}$ (eV) or $\mu$ as illustrated in Figure 4(c). This data is extracted from the surface plot shown in Figure 4(a).

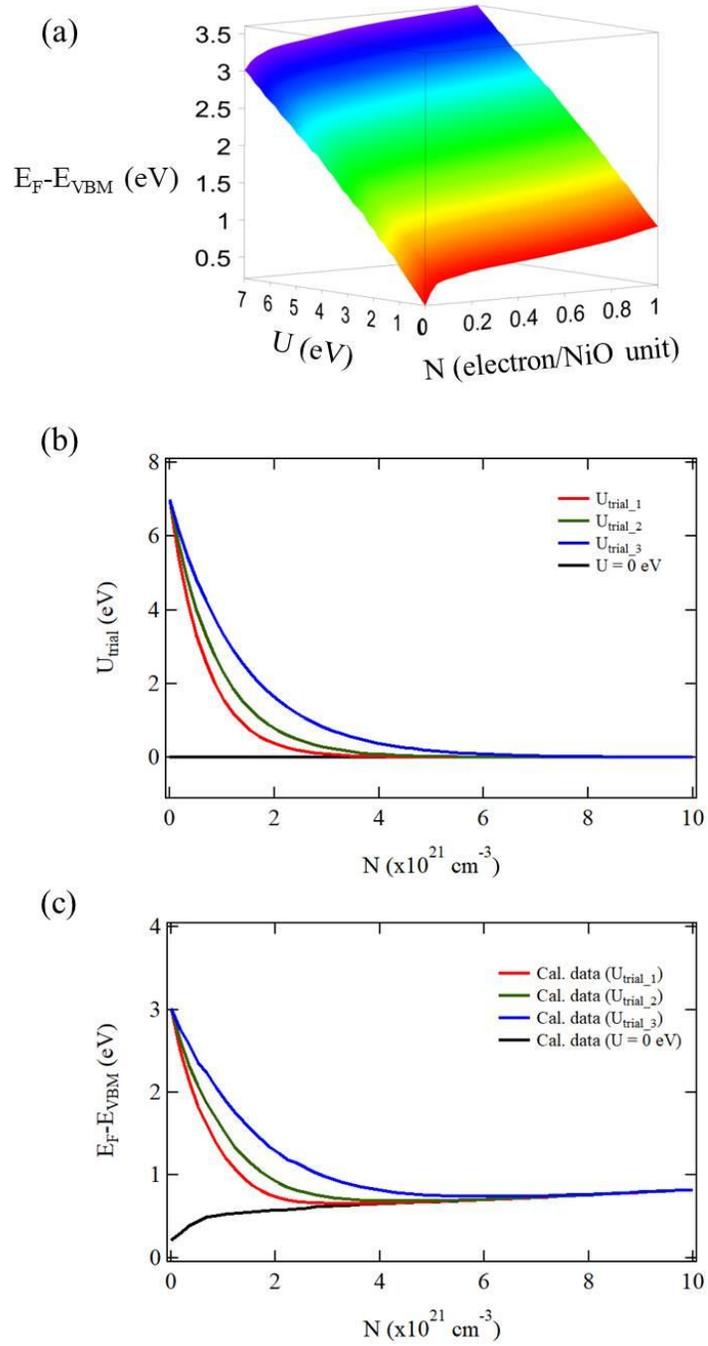

**FIG. 4.** (a) The surface plotting of $E_F$-$E_{VBM}$ correlates with the Hubbard potential (U) and the added electron number (N), as examined through DFT+U calculations. (b) The simplified trial function represents the evolution of Hubbard U as a function of electron. Here, it is simply represented by the exponential formula $U(N) = U_0 \exp(-\alpha N)$. (c) The dynamics of the chemical

potential in strongly correlated NiO show a decrease in the chemical potential during the initial doping before a normal increase at some threshold doping concentration.

Upon the Hubbard potential ($U_{trial}$) decreasing as a function of the charge carrier doping, as shown in Figure 4(b), it can cause the chemical potential (μ) of NiO to decrease for the initially electron-doping condition. This behavior is initiated by the suppression of the correlation gap upon doping, leading to electron compression. At a certain threshold of doping concentration, it is observed that the chemical potential will typically increase, as generally found in a conventional rigid band scenario. From the perspective of doping-dependent U, the DFT investigation convincingly demonstrates the possible occurrence of negative electronic compressibility (NEC). This behavior aligns with experimental findings of NEC in lightly-doped strongly correlated materials $Bi_{0.95}La_{0.05}FeO_3$ (BLFO) and NEC effect in LIB with $BiFe_{0.95}Cu_{0.05}O_3$ coating [17, 46]. Thus, the doping-dependent U has the potential to serve as a generalized model, offering a framework to explain chemical potential evolution in doped Mott-insulators.

In summary, the empirical effect on doped Mott-insulators, given by the decrease in U as a function of charge carrier doping, will lead to the suppression of the Mott-gap and will affect the evolution of the chemical potential. This effect will benefit in the field of energy storage applications, as can be verified in a systematic comparison of capacitance of selected transition metal oxide-based electrodes (Cu, Mn, Co, and Fe oxide electrodes). This finding is supported by a theoretical framework incorporating a model of chemical potential with doping-dependent U. The exploration presents compelling evidence endorsing Mott-insulators as battery electrodes for energy storage. Utilizing their unique properties, Mott-insulators could revolutionize energy storage technology by significantly improving battery capacity.

## SUPPLEMENTARY MATERIAL

The supplementary material contains the tracking data collected from experimental discharged profile of the selected TMOs-based electrode. Besides, the details on estimating the charge storage performance of electrode is provide.

## ACKNOWLEDGMENTS

In memory of J. Zaanen, whose contributions were essential to the core concept of this work. We extend our gratitude to Z. X. Shen for the insightful discussions. This work was financially supported by NSRF via the Program Management Unit for Human Resources & Institutional Development, Research and Innovation (Grant No. B39G670018), JSPS (Grant No. JP22K03535) and NSTC (Grant No. 113WFA0410505). C. Jaisuk acknowledges the Development and Promotion of Science and Technology Talents Project (DPST-scholarship) for financial support. S. Jungthawan was supported by (i) Suranaree University of Technology (SUT), (ii) Thailand Science Research and Innovation (TSRI), (iii) National Science, Research, and Innovation Fund (NSRF) (NRIIS Project Number 195579).

## DATA AVAILABILITY

The data that supports the findings of this study are available from the corresponding author upon reasonable request.